\documentclass[showpacs,preprintnumbers]{revtex4}
\usepackage{amsmath}
\usepackage{graphicx}

\begin{document}

\textbf{Comment on ``R\"{o}ntgen Quantum Phase Shift: A Semiclassical}

\textbf{Local Electrodynamical Effect?''}\bigskip

Tomislav Ivezi\'{c}

\textit{Ru%
\mbox
{\it{d}\hspace{-.15em}\rule[1.25ex]{.2em}{.04ex}\hspace{-.05em}}er Bo\v
{s}kovi\'{c} Institute, P.O.B. 180, 10002 Zagreb, Croatia}

\textit{ivezic@irb.hr\bigskip }

Recently Horsley and Babiker [1] reported semiclassical calculations of the
R\"{o}ntgen and the Aharonov-Casher phase shifts. Their derivation shed
light on a quantum phenomenon without using full apparatus of quantum
theory. They argued that the R\"{o}ntgen phase shift is due to the action of
the nonvanishing three-dimensional (3D) force $\mathbf{F}=\mathbf{\nabla (d}%
\cdot \mathbf{E})$, where $\mathbf{E}=\mathbf{v}\times \mathbf{B}$. Here,
the main conclusion is as in [1], i.e., that the mentioned phase shifts can
be derived using a semiclassical calculation. However, instead of the 3D
force $\mathbf{F}$ from [1] the Lorentz invariant expression for the 4D
force $K^{a}$ is introduced.

First, in [1], $\mathbf{F}$ is written in the dipole rest frame $S^{\prime }$%
, since the electric field (it is $\mathbf{E}^{\prime }$) is obtained by
means of the usual transformations (UT) of $\mathbf{B}$ (e.g., [2], Eq.
(11.149)) from the laboratory frame $S$ to $S^{\prime }$, and $\mathbf{d}$
from [1] is also in $S^{\prime }$, i.e., it is $\mathbf{d}^{\prime }$. For
comparison with experiments the force has to be in $S$ and not in $S^{\prime
}$. Furthermore, it is proved in [3] that the UT of $\mathbf{E}$, $\mathbf{B}
$ are not the Lorentz transformations (LT) and that they have to be replaced
by the LT of the corresponding 4D geometric quantities.

Here, as in [4], we deal with 4D geometric quantities that are defined
without reference frames. They will be called the absolute quantities (AQs).
The notation is the same as in [4]. Different covariant expressions for a
charged particle with a dipole moment are recently presented in [5]. (Here
we consider an uncharged particle.) They include, e.g., the Lagrangian and
the equations of motion for 4-position. However the usual covariant
formulation deals with components, which are implicitly taken in some basis,
mainly the standard basis \{$e_{\mu };\ 0,1,2,3$\} of orthonormal 4-vectors
with $e_{0}$ in the forward light cone. In [4] we have already generalized
the interaction term from the Lagrangian [5] and expressed it in terms of
AQs as $-(1/2)F_{ab}D^{ab}$. Also, in [4], $D^{ab}$ and $F^{ab}$ are
decomposed and written in terms of AQs, the electric and magnetic dipole
moments $d^{a}$ and $m^{a}$ and 4-vectors $E^{a}$ and $B^{a}$. Here the
equation of motion for the 4-position, [5] Eq. (11), is similarly
generalized as $m\overset{..}{x^{c}}=K^{c}=(1/2)D^{ab}\partial ^{c}F_{ab}$
(the dot means the derivation with respect to the proper time $s$). This 4D
force $K^{a}$ replaces the forces written with the 3D $\mathbf{d}$, $\mathbf{%
m}$ and $\mathbf{E}$, $\mathbf{B}$ from [1] and many others.

For comparison with the usual formulation with the 3-vectors one needs to
write the representation of the AQ $K^{a}$ in the $\{e_{\mu }\}$ basis as $%
K^{a}=K^{\alpha }e_{\alpha }=((1/2)D^{\mu \nu }\partial ^{\alpha }F_{\mu \nu
})e_{\alpha }$. Then, [4], we also need to choose the frame of ``fiducial''
observers in which the observers who measure $E^{a}$, $B^{a}$ are at rest.
That frame with the $\{e_{\mu }\}$ basis will be called the $e_{0}$-frame.
In the $e_{0}$-frame the velocity of ``fiducial'' observers $v^{a}=ce_{0}$
and $E^{0}=B^{0}=0$. In that frame $K^{\alpha }$ is the sum of two terms,
one with the electric dipole $(1/c^{2})[cd_{0}(u_{\nu }\partial ^{\alpha
}E^{\nu })-c^{2}(d_{\nu }\partial ^{\alpha }E^{\nu })+c^{2}\varepsilon
^{0ijk}u_{j}d_{k}\partial ^{\alpha }B_{i}]$ (responsible for the R\"{o}ntgen
effect) and another one (responsible for the Aharonov-Casher effect) in
which $d^{\mu }$ is replaced by $m^{\mu }$, while $E^{\mu }$ and $B^{\mu }$
are interchanged . The last term in the above square braces describes the
direct action of the magnetic field on the electric dipole moment. This is
very important difference relative to all previous treatments, e.g., [1].

For comparison with experiments the laboratory frame has to be taken as the $%
e_{0}$-frame. Then, for the R\"{o}ntgen effect from [1], $m^{\mu }=0$, $%
E^{\mu }=0$, $B^{\mu }=(0,B^{1},B^{2},0)$ and $u^{\mu
}=(u^{0},u^{1},u^{2},0) $, since in $S^{\prime }$ $u^{\prime \mu }=(c,0,0,0)$
($u^{a}=dx^{a}/ds$ is the 4-velocity of the particle). $K^{\alpha }$ becomes
$K^{\alpha }=\varepsilon ^{0ijk}u_{j}d_{k}\partial ^{\alpha }B_{i}$, or
explicitly, $K^{0}=K^{3}=0$, $K^{1,2}=d_{3}[u_{2}\partial
^{1,2}B_{1}-u_{1}\partial ^{1,2}B_{2}]$. This force is in $S$ and not in $%
S^{\prime }$ as in [1]. Also $K^{\alpha }$ is not zero even in the case when
the electric dipole moment is aligned parallel to the magnetic line charge.
Only in the case when $S^{\prime }$ is chosen to be the $e_{0}$-frame then
it will be $K^{\alpha }=0 $, but that case is not physically realizable. The
second term in $K^{\alpha }$ gives the analogous result for the
Aharonov-Casher effect. Hence, as in [1], both phase shifts could be
calculated using the concept of force, but not the 3D force than the 4D
force $K^{a}$. Then the phase shifts can be calculated, e.g., as in [6].
\bigskip

\noindent \textbf{References\medskip }

\noindent \lbrack 1] S.A.R. Horsley and M. Babiker, Phys. Rev. Lett. \textbf{%
95}, 010405 (2005).

\noindent \lbrack 2] J.D. Jackson, \textit{Classical Electrodynamics} 3rd
ed. (Wiley, New York, 1998).

\noindent \lbrack 3] T. Ivezi\'{c}, Found. Phys. \textbf{33}, 1339 (2003)%
\textbf{; }Found. Phys. Lett. \textbf{18,} 301 (2005); Found.

Phys. \textbf{35,} 1585 (2005).

\noindent \lbrack 4] T. Ivezi\'{c}, Phys. Rev. Lett. \textbf{98}, 108901
(2007).

\noindent \lbrack 5] A. Peletminskii, S. Peletminskii, Eur. Phys. J. C
\textbf{42}, 505 (2005).

\noindent \lbrack 6] J. Anandan, Int. J. Theor. Phys. \textbf{19}, 537
(1980).

\end{document}